%% file: mtns2020_accepted.tex
\documentclass{article}
\usepackage[utf8]{inputenc}
\usepackage{graphicx}      
\usepackage{amsmath,amssymb,bbm,enumerate}
\usepackage{xcolor}

\input{definitions}

\newtheorem{assumption}{Assumption}
\newtheorem{remark}{Remark}

\newtheorem{lemma}{Lemma}
\newtheorem{theorem}{Theorem}

\newcommand{\python}{\textsc{Python}}
\newcommand{\ssd}{\textit{ssd}}

\newcommand{\tc}{\textcolor}

\newenvironment{pf}{\textbf{Proof}}{\hfill$\Box$}

\title{Robust nonlinear observer design based on impulsive dissipativity\footnote{Manuscript accepted for publication in the proceedings of the IFAC 24th International Symposium on Mathematical Theory of Networks and Systems (MTNS 2020): Cambridge, UK to take place in 2021}}

\author{A. Schaum$^a$, P. Feketa$^a$, T. Meurer$^a$, J. A. Moreno$^b$\\
$^a$ Chair of Automatic Control, \\
Kiel University, 24148 Kiel, Germany \\
(e-mail: \{alsc,pf,tm\}@tf.uni-kiel.de).
\\
$^b$ Instituto de Ingenieria, Universidad Nacional Autonoma de Mexico (UNAM)\\ 
Coyoacan, Mexico City, Mexico \\
(e-mail: \{jmorenop@ii.unam.mx)}

\date{}

\begin{document}

\maketitle

\begin{abstract}
The paper considers the design of a nonlinear dissipative impulsive observer based on non-periodic discrete-time measurements. Sufficient conditions \tc{black} are derived for \tc{black}{(i)} exponential convergence of the observer in absence of measurement uncertainty, and \tc{black}{(ii)} input-to-state stability (ISS) with respect to measurement uncertainty, by combining notions from impulsive and dissipative systems theory. The conditions mainly include constraints on the minimum and maximum time between measurements \tc{black}{depending on} system characteristics, the correction gain and the desired ISS gain. A representative case example is used to illustrate the theoretical assessments.
\end{abstract}

\textbf{keywords}
 Dissipativity, impulsive systems, robustness, nonlinear observer design, measurement uncertainty

\section{Introduction}

Sampled data observer design for nonlinear systems has been extensively studied over the past 50 years. The typical scenario consists in periodic or non-periodic discrete-time measurements for continuous dynamic systems like for the classical continuous-discrete Kalman Filter \cite{Gelb}. A common approach consists in emulating a continuous-time measurement observer using properly chosen time-varying correction terms \cite{Ahmed-AliEtAl2016,Deza1992,Nadri2003}. 
A \tc{black}{recent} study on this approach for nonlinear systems is presented in \cite{Karafyllis2019_arxiv} where some of the previous approaches are generalized and put in the context of input-to-state stability (ISS) \textcolor{black}{\cite{Sontag1989}} with respect to the measurement uncertainty. Besides these approaches also continuous-discrete interval observers in \cite{Mazenc2013,Mazenc2014}, and moving horizon estimation techniques \cite{RawlingsDiehl2016} have been proposed to address the problem. An alternative but similar problem set-up consists in so-called Lebesgue measuring, where instead of an analog value a digital flag is set when some state reaches a certain threshold. For linear systems with Lebesgue measurement a thorough analysis is provided e.g. in \cite{MorenoBetancur2010}.
\tc{black}{Note that the observer in \cite{MorenoBetancur2010} is impulsive in nature and achieves finite-time convergence in the absence of errors. In contrast, the approach e.g. in \cite{Karafyllis2019_arxiv} is based on recovering the behavior of an asymptotically convergent continuous-time observer.}

If instead of continuous state correction only instantaneous measurement injection takes place the observer error dynamics is an impulsive dynamical system. Impulsive observer design has recently \textcolor{black}{been} analyzed in \cite{Feketa2019_cdc} for a class of bioreactor models, where ISS with respect to the measurement uncertainty was ensured if a maximum time between samples is maintained. In this work the observation error dynamics consists of a cascade of an asymptotically stable continuous system and an impulsive system, which was rendered ISS by the impulsive correction. The problem was solved using ISS Lyapunov functions, see e.g., \cite{SontagWang1995,HLT08,DM13,FB19}.

In the present work the impulsive observer design in \cite{Feketa2019_cdc} is extended towards a class of nonlinear systems with partial state measurements leading to an interconnected error dynamics with impulsive innovation. The core problem in comparison to \cite{Feketa2019_cdc} resides in the fact that here no cascade structure can be exploited but the maximum time between samples must be adjusted so that the interconnection is exponentially stable in the absence of measurement uncertainty. Furthermore, a condition on a minimum time between measurements is established to ensure ISS with respect to the measurement uncertainty with a prescribed gain. The problem is addressed within the framework of dissipative dynamical systems \cite{Willems,Haddad2001a}, in particular extending the dissipativity-based observer design for nonlinear systems with continuous measurements \cite{Moreno2004Oaxaca,Schaum2006,Moreno2008} to the case of irregularly sampled measurements.


\section{Problem formulation}

\subsection{Observation problem}
Consider a nonlinear system of the form
\begin{subequations}\label{system}
 \begin{align}
  \dot\bx &= A\bx+\bpsi(\bx),& 0<t\notin \mathbb T\\
  \bs y_k &= C\bx(t_k)+\bw_k,& \textcolor{black}{k\in\mathbb N}
 \end{align}
\end{subequations}
with \textcolor{black}{$\bx:[0,\infty)\to\mathbb R^n$}, $\bx(0)=\bx_0$, $\bpsi\in\mathcal C^1$ a Lipschitz continuous function, output $\bs y(t)\in\mathbb R^m$ according to the measurement matrix $C\in\mathbb R^{m\times n}$ with rank$(C)=m<n$, $\mathbb T=\{t_1,\ldots\}$ the set of sampling instants with $\lim_{k\to\infty}t_k=\infty$ and $\bw_k$ a time-varying measurement uncertainty that is not necessarily gaussian. Additional assumptions on the matrices and functions will be stated in Section \ref{Sect:ObsDesign}.

The problem considered here consists in designing an observer with state estimate $\hbx$ based on the sampled measurements so that the estimation error $\be=\hbx-\bx$ is ISS with respect to the measurement uncertainty, i.e. there exist functions $\alpha\in \mathcal{KL}$ \textcolor{black}{(positive, monotonically increasing/decreasing)} and $\beta\in\mathcal K$ \textcolor{black}{(positive, monotonically increasing)} such that 
\begin{align}\label{def:iss}
 \|\be(t)\|\leq \alpha\left(\|\be(0)\|,t\right)+\beta\left(\|\bs w\|_\infty\right) \quad\textcolor{black}{\forall\, t\geq0}
\end{align}
with $\|\bs w\|_\infty=\sup_{k\in\mathbb N}\|\bs w_k\|$ denoting the infinity-norm.


\subsection{Notions from dissipativity theory}

Following the notion in \cite{Willems,Willems2} and further discussed in \cite{HillMoylan1980} a system with state $\bx$, input $\bs u$ and output $\bs y$ is called dissipative with respect to a supply rate $\omega(\bs y,\bs u)$ if there exists a storage function $\mathcal S(\bx)\geq0$ so that
\begin{align*}
 \mathcal S(\bx(t))\leq \mathcal S(\bx(0))+\int_0^t\omega(\by(\tau),\bu(\tau))\dtau,
\end{align*}
or if $\mathcal S$ is differentiable
\begin{align*}
 \ddt{\mathcal S} &\leq \omega(\by,\bu).
\end{align*}
In this case the system is called strictly state dissipative with dissipation rate $\kappa$, or for short \ssd($\kappa$), if 
\begin{align}\label{def:ssd}
 \ddt{\mathcal S} &\leq -\kappa\|\bx\|^2 + \omega(\by,\bu).
\end{align}
For a quadratic supply rate 
\begin{align}\label{def:qsr}
 \omega(\by,\bu) =  \begin{bmatrix}\by\\ \bu \end{bmatrix}^\intercal
                    \begin{bmatrix} Q & S\\ S^\intercal & R\end{bmatrix}
                    \begin{bmatrix}\by\\ \bu \end{bmatrix}\geq0
\end{align}
the system is called $(Q,S,R)$-\ssd($\kappa$) if \eqref{def:ssd} holds true with $\omega$ given by \eqref{def:qsr}.

In the following, $\Sigma(A,B,C)$ denotes a linear system of the form
\begin{subequations}
\begin{align}
 \dot \bx&=A \bx+B\bs u\\
 \bs y &= C\bx
\end{align}
\end{subequations}
with vectors and states of appropriate dimension. Considering the quadratic storage function
\begin{align}\label{QuadStor}
 \mathcal S = \bx^\intercal\bx
\end{align}
it follows that $\Sigma(A,B,C)$ is $(Q,S,R)$-\ssd($\kappa$) if
\begin{align}\label{cond:qsr}
 \begin{bmatrix} A+A^\intercal+\kappa I & B\\ B^\intercal & 0\end{bmatrix}\leq \begin{bmatrix} Q & S\\ S^\intercal & R\end{bmatrix}.
\end{align}

A static, memoryless map $\bs \varphi(\bu)$ with $\bs\varphi(\bs 0)=\bs 0$ is called $(Q,S,R)$-dissipative if the associated supply rate in \eqref{def:qsr} is non-negative, i.e. 
\begin{align}\label{def:qsr_stat}
\omega(\bs\varphi,\bu)=
	\begin{bmatrix} \bs\varphi \\ \bs\nu \end{bmatrix}^\intercal
	\begin{bmatrix} Q & S \\ S^\intercal & R \end{bmatrix}
	\begin{bmatrix} \bs\varphi \\ \bs\nu \end{bmatrix}
	\geq0.
\end{align}

\section{Impulsive dissipativity-based observer}\label{Sect:ObsDesign}

Consider the following state transformation
\begin{align}
 \bz_o &= C\bx,\quad \bz_n = M\bx,\quad T=\begin{bmatrix}C\\ M\end{bmatrix},\quad \det\left(T\right)\neq 0,
\end{align}
with $M\in\mathbb R^{(n-m)\times n}$. 
Note that by the condition rank$(C)=m$ and $C\in\mathbb R^{m\times n}$ this transformation always exists. The dynamics in the $\bz=[\bz_o^\intercal~\bz_n^\intercal]^\intercal$ coordinates reads
\begin{subequations}
\begin{align}
 \dot\bz &= \bar A\bz + \bar\bpsi(\bz),& t\notin\mathbb T\\
 \by_k &= \bz_o\textcolor{black}{(t_k) + \bs w_k = \bar C\bz(t_k) + \bs w_k},& \textcolor{black}{k\in\mathbb N}
\end{align}
with the matrices
\begin{align}
 \bar A &= TAT^{-1} = \begin{bmatrix}
             \bar A_{o} & \bar A_{on} \\ \bar A_{no} & \bar A_{n}
           \end{bmatrix}
 ,\quad \bar C = CT^{-1} = \begin{bmatrix} I & 0\end{bmatrix}
\end{align}
\end{subequations}
\textcolor{black}{and $\bar\bpsi(\bz)=T\bpsi(T^{-1}\bz)$}.

The observer is proposed as follows:
\begin{subequations}\label{observer}
 \begin{align}
  \dot\hbz &= \bar A\hbz+\bar\bpsi(\hbz),& 0<t\notin \mathbb T\\
  \hbz_o^+(t_k) &= \hbz_o(t_k) - L\left(\hbz_o(t_k)-\by_k\right),& \textcolor{black}{k\in\mathbb N}
 \end{align}
\end{subequations}
with diagonal correction gain matrix $L$. \textcolor{black}{In the sequel it is assumed that the state $\hbz(t)$ is left-continuous and that for all $k\in\mathbb N$ there exists the right limit $\hbz^+(t_k)=\lim_{t\downarrow t_k}\hbz(t)$.}

For the purpose of analyzing the observer convergence introduce the observation error
\begin{align}\label{def:error}
 \beps = \hbz-\bz,\quad \beps = \begin{bmatrix}\beps_o\\ \beps_n\end{bmatrix}
\end{align}
with dynamics
\begin{subequations}\label{eq:obsError}
 \begin{align}
  \dot\beps &= \bar A\beps+\tilde\bpsi(\beps),& 0<t\notin \mathbb T\\
  \beps_o^+(t_k) &= (I- L)\beps_o(t_k) + L\bs w_k,& \textcolor{black}{k\in\mathbb N}
 \end{align}
and 
\begin{align}
 \tilde\bpsi(\beps)&=\bar\bpsi(\bz+\beps)-\bar\bpsi(\bz).
\end{align}
\end{subequations}

\begin{figure}[!h]
 \begin{picture}(2,4.65)
  \put(0.5,0){\includegraphics[scale=0.5]{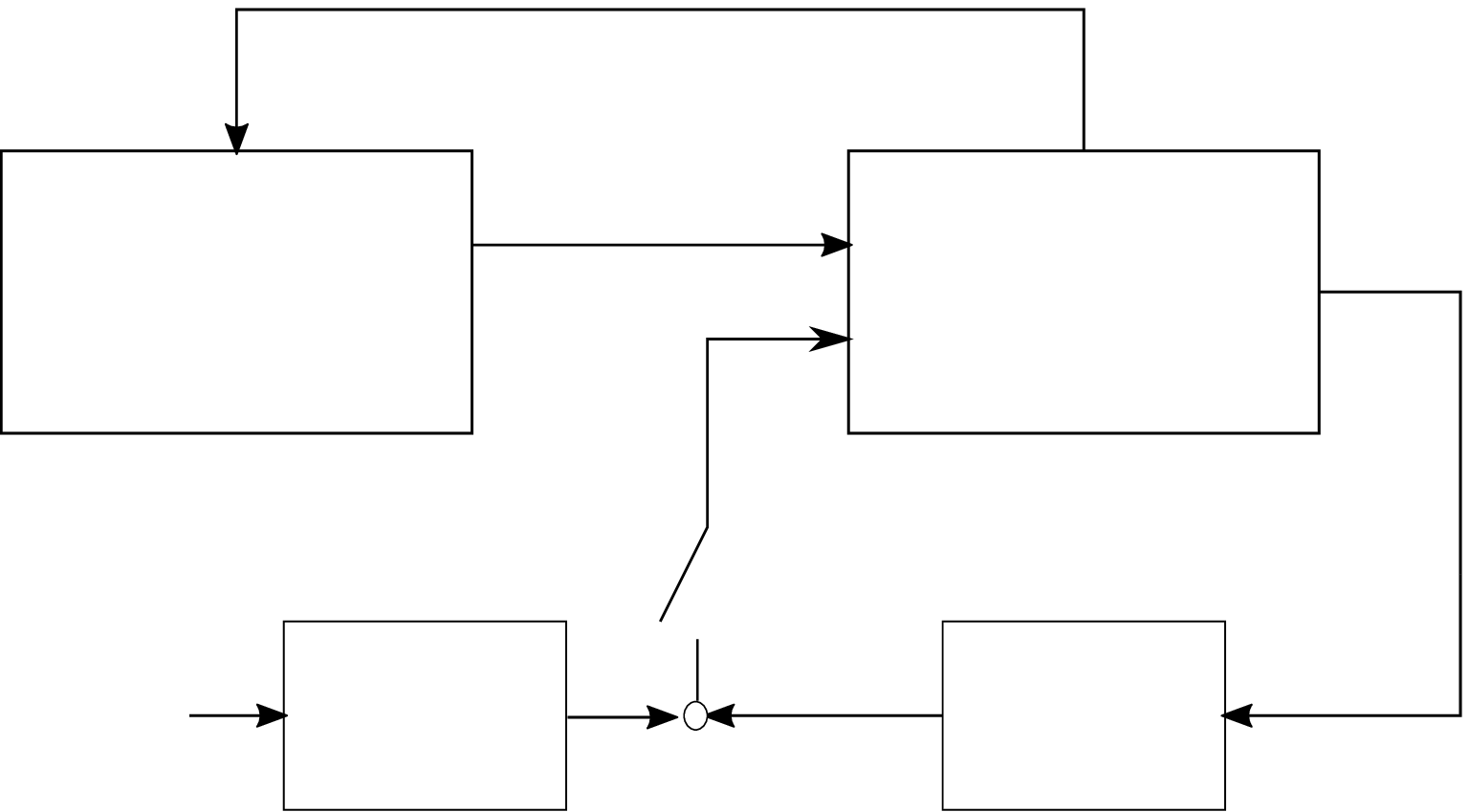}}
  \put(1.45,2.7){\Large $\Sigma_n$}
  \put(6,2.7){\Large $\Sigma_o$}
  \put(5.8,0.5){\large $I-L$}
  \put(2.6,0.5){\large $L$}
  \put(1.1,0.5){\large $\bs w_k$}
  \put(4.5,1.35){$\textcolor{black}{t_k\in\mathbb T}$}
 \end{picture}
 \caption{Interconnection of the continuous and impulsive dynamical system given in \eqref{errorDynamics}.}
 \label{Fig:SysInterconnection}
\end{figure}

The preceding dynamics corresponds to an interconnection of \textcolor{black}{the impulsive} system \textcolor{black}{$\Sigma_o$} and the \textcolor{black}{non-impulsive one} \textcolor{black}{$\Sigma_n$}
\begin{subequations}\label{errorDynamics}
 \begin{align}
 \label{errorDynamics_o}
  \Sigma_o:&\left\{\begin{array}{cc}
  ~~~~~\dot\beps_o = \bar A_{o}\beps_o + \bar A_{on}\beps_n + \tilde\bpsi_o(\beps),& 0<t\notin \mathbb T\\
  \beps_o^+(t_k) = (I- L)\beps_o(t_k)+L\bs w_k,& \textcolor{black}{k\in\mathbb N}
  \end{array}\right.\\[1ex]
  \label{errorDynamics_n}
  \Sigma_n:&\quad ~
  \dot\beps_n = \bar A_{no}\beps_o + \bar A_{n}\beps_n + \tilde\bpsi_n(\beps), t\geq0.
 \end{align}
\end{subequations}
This interconnection is schematically depicted in Figure \ref{Fig:SysInterconnection} where $\Sigma_o$ stands for the hybrid system \eqref{errorDynamics_o} with innovation by measurement injection, and $\Sigma_n$ for the continuous dynamics \eqref{errorDynamics_n} without innovation.
Note that in comparison to the study in \cite{Feketa2019_cdc} here an additional interconnection between the continuous and impulsive dynamics has to be considered. 

In the sequel the following assumptions on the functions $\tilde\bpsi_i(\beps),\, i\in\{o,n\}$ are considered.
\begin{assumption}
For $i\neq j\in\{o,n\}$ the map $\tilde\bpsi_i(\beps)$ is $(Q_i,S_i,R_i)$ dissipative with $Q_i\preceq0$, uniformly with respect to $\beps_j$, i.e. for all $\beps_j$ it holds that
\begin{align}
 \begin{bmatrix}\tilde\bpsi_i\\\beps_i\end{bmatrix}^\intercal
 \begin{bmatrix}Q_i & S_i \\ S_i^\intercal & R_i\end{bmatrix}
 \begin{bmatrix}\tilde\bpsi_i\\\beps_i\end{bmatrix}
 \geq 0.
\end{align}
\end{assumption}
\vspace{0.25cm}

The following result for the dynamics \eqref{errorDynamics_n} of the unmeasured state $\beps_n$ is crucial in the following analysis.
\begin{lemma}\label{lemma:dissip_Sigma_n}
 Let Assumption 1 hold true and let $\Sigma(\bar A_{n},I,I)$ be $(-R_n,-S_n^\intercal,-Q_n)$-\ssd($\kappa_n$). Then system \eqref{errorDynamics_n} with input $\beps_o$ and output $\beps_n$ is \ssd($\kappa_n$) with respect to the supply rate $\omega_n(\beps_n,\beps_o)=2\beps_n^\intercal \bar A_{no}\beps_o$ and the storage function $\mathcal S_n=\|\beps_n\|^2\succ0$. If in addition \tc{black}{$\kappa_n>\lambda_{no}^*=\|\bar A_{no}\|$}, then the system \eqref{errorDynamics_n} is ISS with respect to $\beps_o$.
\end{lemma}
\begin{pf}
The rate of change of the storage function $\mathcal S_n(\beps_n)$ is given by
\begin{align*}
 \ddt{\mathcal S_n} &= \beps_n^\intercal \left(\bar A_{n}^\intercal+\bar A_{n}\right)\beps_n + 2\beps_n^\intercal A_{no}\beps_o + 2\beps_n^\intercal\tilde\bpsi_n(\beps)\\
           &= \begin{bmatrix}\beps_n\\ \tilde\bpsi_n\end{bmatrix}^\intercal
              \begin{bmatrix}\bar A_{n}^\intercal+\bar A_{n} & I \\ I & 0\end{bmatrix}
              \begin{bmatrix}\beps_n\\ \tilde\bpsi_i\end{bmatrix} + 2\beps_n^\intercal \bar A_{no}\beps_o
              \\
           &\leq -\kappa_n\|\beps_n\|^2 + 
                \begin{bmatrix}\beps_n\\ \tilde\bpsi_n\end{bmatrix}^\intercal
                \begin{bmatrix}-R_i & -S_i^\intercal \\ -S_i & -R_i\end{bmatrix}
                \begin{bmatrix}\beps_n\\ \tilde\bpsi_i\end{bmatrix} 
                + 2\beps_n^\intercal \bar A_{no}\beps_o
                \\
           &\leq -\kappa_n\|\beps_n\|^2 -
                \begin{bmatrix}\tilde\bpsi_i\\\beps_i\end{bmatrix}^\intercal
                \begin{bmatrix}Q_i & S_i \\ S_i^\intercal & R_i\end{bmatrix}
                \begin{bmatrix}\tilde\bpsi_i\\\beps_i\end{bmatrix}
                + 2\beps_n^\intercal \bar A_{no}\beps_o
                \\
           &\leq -\kappa_n\|\beps_n\|^2 +\omega_n(\beps_n,\beps_o).
\end{align*}
This shows that \eqref{errorDynamics_n} is \ssd($\kappa_n$) with respect to the supply rate $\omega_n$. Furthermore, \tc{black}{with $\lambda_{no}^*=\|\bar A_{no}\|$} it holds that
\begin{align*}
 2\beps_n^\intercal\bar A_{no}\beps_o \leq 2\lambda_{no}^*\|\beps_n\|\,\|\beps_o\|\leq \lambda_{no}^*\left(\|\beps_n\|^2+\|\beps_o\|^2\right)
\end{align*}
and thus
\begin{align*}
 \ddt{\mathcal S_n} &\leq -\left(\kappa_n-\lambda_{no}^*\right)\|\beps_n\|^2 + \lambda_{no}^*\|\beps_o\|^2.
\end{align*}
Thus, for $\kappa_n>\lambda_{no}^*$ the function $\mathcal S_n$ is an ISS-Lyapunov function \cite{SontagWang1995}.
\hfill$\Box$
\end{pf}

The following analysis is divided into two steps: first sufficient conditions on the maximum interval between measurements for which the convergence of the observer is ensured are derived for the case that $\bs w_k=\bs 0$ for all $k\in\mathbb N$. Secondly, sufficient conditions for the ISS of the observation error with respect to the measurement uncertainty with a prescribed gain are established in terms of the minimum interval (dwell-time) between measurements.

\subsection{Convergence in absence of measurement uncertainty}

\begin{lemma}\label{lemma:aux_eps2}
 Let Assumption 1 hold true \tc{black}{with $Q_o\prec0$}. Then for
 \begin{align}\label{auxCond_varpi_o}
  \varpi_o>\max_{\lambda\in\sigma(M_o)}\lambda
 \end{align}
 with \tc{black}{$M=\bar A_{o}+\bar A_{o}^\intercal+R_o-(I+S_o^\intercal)Q_o^{-1}(I+S_o)$} the system $\Sigma(\bar A_{o},I,I)$ is \textcolor{black}{$(-R_o,-S_o^\intercal,-Q_o)$}-\ssd($-\varpi_o$) and the system \eqref{errorDynamics_o} with input $\beps_n$ and output $\beps_o$ is piecewise \ssd($-\varpi_o$) with respect to the supply rate $\omega_o=2\beps_o^\intercal\bar A_{on}\beps_n$.
\end{lemma}
\begin{pf}
 By definition $\Sigma(\bar A_{o},I,I)$ is \textcolor{black}{$(-R_o,-S_o^\intercal,-Q_o)$} \ssd(-$\varpi_o$) if 
 \begin{align*}
 \begin{bmatrix} \bar A_{o}+\bar A_{o}^\intercal+R_o-\varpi_o I  & I + S_o^\intercal \\ I+S_o & Q_o\end{bmatrix}\preceq 0.
\end{align*}
Since by assumption \tc{black}{$Q_o\prec0$} it follows from the Schur complement \cite{Dym} that this holds true if
\begin{align*}
 \tc{black}{\bar A_{o}+\bar A_{o}^\intercal+R_o-\varpi_oI-(I+S_o^\intercal)Q_o^{-1}(I+S_o)\prec 0},
\end{align*}
or equivalently if \eqref{auxCond_varpi_o} is satisfied. Having ensured this, the piecewise dissipativity property of \eqref{errorDynamics_o} with respect to $\omega_o$ follows as in the proof of Lemma \ref{lemma:dissip_Sigma_n}.
 \hfill$\Box$
\end{pf}
\begin{remark}\label{remark:varpi_o_largerZero}
  Note that $\varpi_o\in\mathbb R$ in Lemma \ref{lemma:aux_eps2} can \textcolor{black}{be positive and thus} \eqref{auxCond_varpi_o} \textcolor{black}{can always be} satisfied. \textcolor{black}{Nevertheless, $\varpi_o>0$ corresponds to} a negative dissipation. Thus, in contrast to Lemma \ref{lemma:dissip_Sigma_n} for the continuous part of $\Sigma_o$ no ISS property with respect to $\beps_n$ can be established at this point.
\end{remark}

\begin{assumption}\label{ass:varpi}
 Motivated by Remark \ref{remark:varpi_o_largerZero} in the sequel it is assumed that \eqref{auxCond_varpi_o} is satisfied with $\varpi_o>0$.
\end{assumption}

Consider the storage function 
\begin{align*}
 \mathcal S_o(\beps_o) &= \beps_o^\intercal \beps_o = \|\beps_o\|^2
\end{align*}
Note that with Assumption \ref{ass:varpi} the continuous evolution of $\mathcal S_o$ along the solution of \eqref{errorDynamics_o} is increasing with time. Anyway, during jumps the error norm $\|\beps_o\|$ will be reduced significantly according to \eqref{errorDynamics_o} and thus
\begin{align}
 \mathcal S_o(\beps_o^+(t_k))&=\left[\beps_o(t_k)\right]^\intercal(I-L)^\intercal (I-L)\beps_o(t_k)\\
              &\leq (1-\gamma)^2 \mathcal S_o(\beps_o(t_k))
\end{align}
where $0<(1-\gamma)<1$ is the maximum eigenvalue of the diagonal matrix $I-L$.

\begin{figure}[!h]
 \setlength{\unitlength}{1cm}
 \begin{picture}(2,4.75)
  \put(1,0){\includegraphics[scale=0.6]{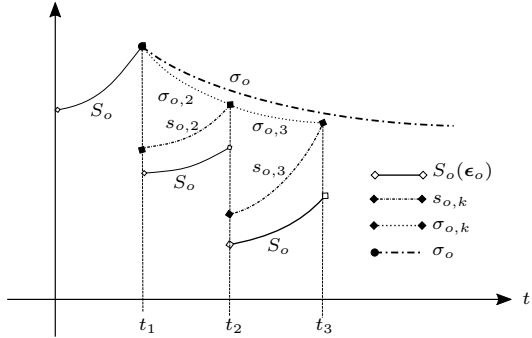}}
  \put(6.65,2.15){\scriptsize $S_o(\beps_o)$}
  \put(6.65,1.8){\scriptsize $s_{o,k}$}
  \put(6.65,1.45){\scriptsize $\sigma_{o,k}$}
  \put(6.65,1.1){\scriptsize $\sigma_o$}
  \put(2.1,2.95){\scriptsize $S_o$}
  \put(3.2,2){\scriptsize $S_o$}
  \put(4.45,1.15){\scriptsize $S_o$}
  \put(3.1,2.8){\scriptsize $s_{o,2}$}
  \put(4.25,2.2){\scriptsize $s_{o,3}$}
  \put(3,3.15){\scriptsize $\sigma_{o,2}$}
  \put(4.25,2.75){\scriptsize $\sigma_{o,3}$}
  \put(3.95,3.4){\scriptsize $\sigma_{o}$}
  \put(2.75,0.1){\scriptsize $t_1$}
  \put(3.9,0.1){\scriptsize $t_2$}
  \put(5.1,0.1){\scriptsize $t_3$}
  \put(7.85,0.45){\scriptsize $t$}
 \end{picture}
 \caption{Construction of the bounding function $s_{o,k},\sigma_{o,k}$ and $\sigma_o$ for a possible evolution of $S_o(\beps_o)$.}
 \label{Fig:construction}
\end{figure}

Accordingly, if it is ensured that the maximum increase of $\mathcal S_o$ between jumps is \tc{black}{smaller than} the reduction during jumps, an overall error dissipation can be established and a continuous and differentiable function $\sigma_o$ can be constructed that bounds $\mathcal S_o(\beps_o)$ for all $t\geq t_1$. The construction of this function is addressed next, and the different steps are illustrated in Figure \ref{Fig:construction}. 

The rate of change of $\mathcal S_o$ over the continuous solution parts of \eqref{errorDynamics_o} is given by
\begin{align*}
 \ddt{\mathcal S_o} = \begin{bmatrix}\beps_o\\ \tilde\bpsi_o\end{bmatrix}^\intercal
                \begin{bmatrix} \bar A_{o}+\bar A_{o}^\intercal  & I \\ I & 0\end{bmatrix}
                \begin{bmatrix}\beps_o\\ \tilde\bpsi_o\end{bmatrix} + 2\beps_o^\intercal \bar A_{on}\beps_n
\end{align*}
Under the assumptions of Lemma \ref{lemma:aux_eps2} it follows that
\begin{align*}
 \ddt{\mathcal S_o} &\leq \varpi_o \|\beps_o\|^2 + 2\beps_o^\intercal \bar A_{on}\beps_n.
\end{align*}
Accordingly, for $t\in[t_{k-1},t_k)$ it holds that
\begin{align*}
 \mathcal S_o(\beps_o(t)) &\leq e^{\varpi_o (t-t_{k-1})}\mathcal S_o(\beps_o^+(t_{k-1})) \\
        &\hspace{2cm}+ 2\int_{t_{k-1}}^te^{\varpi_o (t-\tau)} \beps_o^\intercal(\tau) \bar A_{on}\beps_n(\tau)\dtau\\
        &\leq (1-\gamma)^2 e^{\varpi_o (t-t_{k-1})} \mathcal S_o(\beps_o(t_{k-1})) \\
        &\hspace{2cm} + 2\int_{t_{k-1}}^te^{\varpi_o (t-\tau)} \beps_o^\intercal(\tau) \bar A_{on}\beps_n(\tau)\dtau\\
        &\leq (1-\gamma)^2 e^{\varpi_o (t-t_{k-1})} \mathcal S_o(\beps_o(t_{k-1})) \\
        &\hspace{1.35cm} + \int_{t_{k-1}}^te^{\varpi_o (t-\tau)} 2\lambda_{on}^*\|\beps_o(\tau)\|\,\|\beps_n(\tau)\|\dtau
\end{align*}
with $\lambda_{on}^*=\|\bar A_{on}\|$. Denote the function on the right-hand side of the preceding inequality as $s_{o,k}$ defined on the interval $[t_{k-1},t_k]$. Using  the variation of constants formula it can be verified that $s_{o,k}$ coincides at $t=t_k$ with the solution of the system
\begin{align*}
 \dot\sigma_{o,k} &= -\kappa_{o,k}\sigma_{o,k} + \beta_k 2\lambda_{on}^*\|\beps_o\|\,\|\beps_n\|
\end{align*}
with initial condition (see Figure \ref{Fig:construction})
\begin{align*}
\sigma_{o,k}(t_{k-1})=\begin{cases} \mathcal S_o(\beps_o(t_{1})), & k=2\\ s_{o,k-1}(t_{k-1}), & \text{else} \end{cases}
\end{align*}
and 
\begin{align*}
 -\kappa_{o,k} &= \frac{\ln((1-\gamma)^2)}{t_k-t_{k-1}}+\varpi_o,\quad 
 \beta_k(t) = \left(\frac{1}{(1-\gamma)^2}\right)^{\frac{t_k-t}{t_k-t_{k-1}}}.
\end{align*}
By construction it holds that $\mathcal S_o(\beps_o(t))\leq \sigma_{o,k}(t)$ for all $t\in[t_{k-1},t_k)$. Note that for 
\begin{align}\label{cond:T}
 T=\max_{k\in\mathbb N} (t_k-t_{k-1})< -\frac{\ln((1-\gamma)^2)}{\varpi_o}
\end{align}
it follows that $\kappa_{o,k}>0$ for all $k\in\mathbb N$ since $0<1-\gamma<1$ holds by assumption. Furthermore, note that $T$ can always be chosen so that \eqref{cond:T} holds true. The relation between the function $\mathcal S_o(\beps_o), s_{o,k}$ and $\sigma_{o,k}$ is illustrated in Figure \ref{Fig:construction}.
Finally, under the condition that $T$ satisfies \eqref{cond:T}, for $t\geq t_1$ it holds that
\begin{align}\label{relation:S_o_sigma_o}
 \mathcal S_o(\beps_o(t))\leq \sigma_o(t),
\end{align}
where $\sigma_o$ is the solution of
\begin{subequations}\label{def:sigma_o}
\begin{align}
 \dot\sigma_o &= -\kappa_o\sigma_o + \beta \lambda_{on}^*\|\beps_o\|\,\|\beps_n\|
\end{align}
with initial condition $\sigma_o(t_1)=\mathcal S_o(\beps_o(t_1))$ and
\begin{align}
 \kappa_o &= \min_{k\in\mathbb N} \kappa_{o,k} = -\left(\frac{\ln((1-\gamma)^2)}{T}+\varpi_o\right)\\
 \beta &= \max_{k\in\mathbb N}\beta_k = \frac{1}{(1-\gamma)^2}.
\end{align}
\end{subequations}
A sketch of a possible evolution of $\sigma_o$ is given in Figure \ref{Fig:construction}.

The function $\sigma_o$ can be used to establish sufficient conditions for the asymptotic convergence of the observer \eqref{observer} as summarized in the next theorem.

\begin{theorem}\label{thm:main}
 Let $\lambda_{on}^*=\|\bar A_{on}\|, \lambda_{no}^*=\|\bar A_{no}\|$. For $\bs w_k=\bs 0$ the observer \eqref{observer} exponentially converges if the conditions of Lemma 1 and 2 hold true, $t_k-t_{k-1}\leq T_{max}$ with $T_{max}$ given in \eqref{cond:T} and for some $\kappa>0$ and for all $i\in\{o,n\}$ it holds that
 \begin{align}\label{condThm1}
  \kappa_i>\lambda_{no}^*+\beta\lambda_{on}^*+\kappa.
 \end{align}
\end{theorem}
\begin{pf}
 According to the preceding construction of the function $\sigma_o$ and given \eqref{relation:S_o_sigma_o}, if $\sigma_o\to 0$ then $\|\beps_0\|\to0$. 
 
 Consider the Lyapunov function candidate given by the overall storage function
 \begin{align}\label{storageFun}
  \mathcal S(\sigma_o,\beps_n)=\sigma_o+\mathcal S_n(\beps_n)\succ0.
 \end{align}
 Considering that condition \eqref{condThm1} holds true it follows that
 \begin{align*}
  \ddt{\mathcal S} 
        &\leq -\kappa_o \sigma_o + 2\beta\lambda_{on}^*\|\beps_o\|\,\|\beps_n\| \\
        &\hspace{1cm} -\kappa_n \|\beps_n\|^2 + 2\lambda_{no}^*\|\beps_o\|\,\|\beps_n\|\\
        &\leq -\kappa_o \sigma_o + \beta\lambda_{on}^*\left(\|\beps_o\|^2+\|\beps_n\|^2\right) \\
        &\hspace{1cm} -\kappa_n \|\beps_n\|^2 + \lambda_{no}^*\left(\|\beps_o\|^2+\|\beps_n\|^2\right)\\
        &\leq -\kappa_o \sigma_o + \beta\lambda_{on}^*\left(\sigma_o+\mathcal S_n\right) \\
        &\hspace{1cm} -\kappa_n \mathcal S_n + \lambda_{no}^*\left(\sigma_o+\mathcal S_n\right)\\
        &\leq -\left(\kappa_o-\lambda_{no}^*-\beta\lambda_{on}*\right) \sigma_o\\
        &\hspace{1cm} -\left(\kappa_n-\lambda_{no}^*-\beta\lambda_{on}*\right) \mathcal S_n\\
        &\leq -\kappa \mathcal S
 \end{align*}
 implying that $\mathcal S(\beps(t))\leq \mathcal S(\beps(0))e^{-\kappa t}$ and thus the exponential convergence of the observer.
 \hfill$\Box$
\end{pf}

After the derivation of sufficient conditions for the exponential convergence of the impulsive observer in absence of measurement uncertainty the effect of the impulsive perturbations introduced by this uncertainty is explicitly considered in the following subsection.

\subsection{Convergence with measurement uncertainty}

Note that in the presence of measurement uncertainty in \eqref{errorDynamics} instead of \eqref{def:sigma_o} one has to consider the impulsive system 
\begin{subequations}\label{def:sigma_o_perturbed}
\begin{align}
 \dot\sigma_o & = -\kappa_o\sigma_o+\beta\lambda_{on}^*\|\beps_o\|\,\|\beps_n\|,&t\notin\mathbb T\\
 \sigma_o^+(t_k) &\leq \sigma_o(t_k)+\gamma \|\bs w_k\|, &\textcolor{black}{k\in\mathbb N}.
\end{align}
For a bounded perturbation the worst increase during jumps is given by
\begin{align}
 \sigma_o^+(t_k) &\leq \sigma_o(t_k) + \gamma \|\bs w\|_\infty, \quad \textcolor{black}{k\in\mathbb N}.
\end{align}
\end{subequations}
Accordingly, the dissipation \eqref{storageFun} of the interconnected system measured in the storage function $\mathcal S$ is given by the impulsive system
\begin{subequations}
\begin{align}
 \dot{\mathcal{S}}(\beps(t))&\leq -\kappa \mathcal S(\beps(t)), &t\notin\mathbb T\\
 \mathcal S^+(\beps(t_k)) &\leq \mathcal  S(\beps(t_k))+\gamma \|\bs w\|_\infty,&\textcolor{black}{k\in\mathbb N}.
\end{align}
\end{subequations}
Given that the measurement uncertainty drives $\mathcal S$ away from the origin, a minimum dwell-time condition between impulses, i.e. a minimum interval between measurements has to be established to ensure that the solution will converge into a positively invariant set of prescribed size $\mathcal S_{max}=\alpha \|\bs w\|_\infty$ with $\alpha > \gamma $.

\begin{theorem}\label{thm:ISS_wm}
Let the conditions of Theorem 1 be satisfied and 
\begin{subequations}\label{cond:T_final}
\begin{align}
 T_{min}< t_{k+1}-t_{k}< T_{max},\quad \textcolor{black}{k\in\mathbb N}
\end{align}
with 
\begin{align}
 T_{max} &= -\frac{\ln\left((1-\gamma)^2\right)}{\varpi_o+\kappa+\lambda_{no}^*+\beta\lambda_{on}^*}\\
 T_{min} &\textcolor{black}{=} -\frac{\ln\left(\frac{\alpha-\gamma}{\alpha}\right)}{\kappa}.
\end{align}
\end{subequations}
Then the observation error dynamics is ISS with respect to the measurement uncertainty with gain $\alpha>\gamma$, and $\beps$ in \eqref{errorDynamics} converges into the positively invariant set 
\begin{align*}
 \mathcal B = \big\{\beps\in\mathbb R^n\,|\, \|\beps\|\leq \sqrt{\alpha \|\bs w\|_\infty}\big\}.
\end{align*}
\end{theorem}
\begin{pf}
To ensure convergence of $\beps$ into the set $\mathcal B$ it is sufficient to show that $\mathcal S(\beps)$ converges into the set $\mathfrak S=\{s\in\mathbb R\,|\, 0\leq s\leq \alpha \|\bs w\|_\infty\}$ and that this set is positively invariant. For the positive invariance it is sufficient that for $\mathcal S(\beps(\textcolor{black}{t_{k-1}}))=\alpha \|\bs w\|_\infty$ it follows that $\mathcal S(\beps(t_k))+\gamma \|\bs w\|_\infty\leq \alpha\|\bs w\|_\infty$, or equivalently
\begin{align}
   \alpha e^{-\kappa (t_k-t_{k-1})}+\gamma\leq \alpha.
\end{align}
This is ensured if $t_k-t_{k-1}$ satisfies $t_k-t_{k-1}>T_{min}$ with $T_{min}$ given in \eqref{cond:T_final}. This condition also ensures that for $\mathcal S(t_{k-1})> \mathcal S_{max}$
\begin{align}\label{finCond1}
  \mathcal  S(t_{k-1})e^{-\kappa (t_k-t_{k-1})}+\gamma \|\bs w\|_\infty < \mathcal S(t_{k-1})
\end{align}
implying the attractivity of the set $\mathfrak S$. \hfill $\Box$
\end{pf}
Note that in addition to the preceding result one can follow the steps outlined in the construction of the function $\sigma_o$ in \eqref{def:sigma_o} to construct a differentiable function $\bar s$ that majorizes $\mathcal S(\beps)$ so that $\|\beps(t)\|\leq \mathcal S(\beps(t))\leq \bar s(t)\,\forall\, t\geq t_1$ and satisfies the ISS relation \eqref{def:iss}
\begin{align}
 \dot{\bar{s}}\leq \bar\alpha(\bar s(t_1),t-t_1)+\bar\gamma(\|\bs w_k\|_\infty)
\end{align}
with a class $\mathcal{KL}$ function $\bar\alpha$ and a class $\mathcal K$ function $\bar\gamma$.

\section{Illustration example}

To illustrate the theoretical observer properties consider 
\begin{subequations}\label{example}
\begin{align}
 \dot x_1 &= \mu(x_1,x_2),&x_1(0)=x_{10}\\
 \dot x_2 &= bx_1+cx_2,&x_2(0)=x_{20}\\
 \mu(x_1,x_2) &= \frac{x_1x_2(x_2+a)(x_2-a)}{1+x_2^4}\\[1ex]
 y_k &= x_1(t_k)+w_k, & \textcolor{black}{k\in\mathbb N}
\end{align}
\end{subequations}
with constants $a,b,c\in\mathbb R$ and $c<0$. Note that due to the non-monotonicity of the function $\mu$ with respect to the unmeasured state $x_2$ it can be easily shown that even with a continuous measurement of $x_1$ system \eqref{example} is not globally observable and additionally it is not locally observable in all extremal points of $\mu(x_1,\cdot)$. Nevertheless, under the condition $c<0$ it is detectable for a continuous measurement of $x_1$. 

\begin{figure}[!h]
 \begin{picture}(2,5.5)
  \put(2,0.25){\includegraphics[scale=0.375]{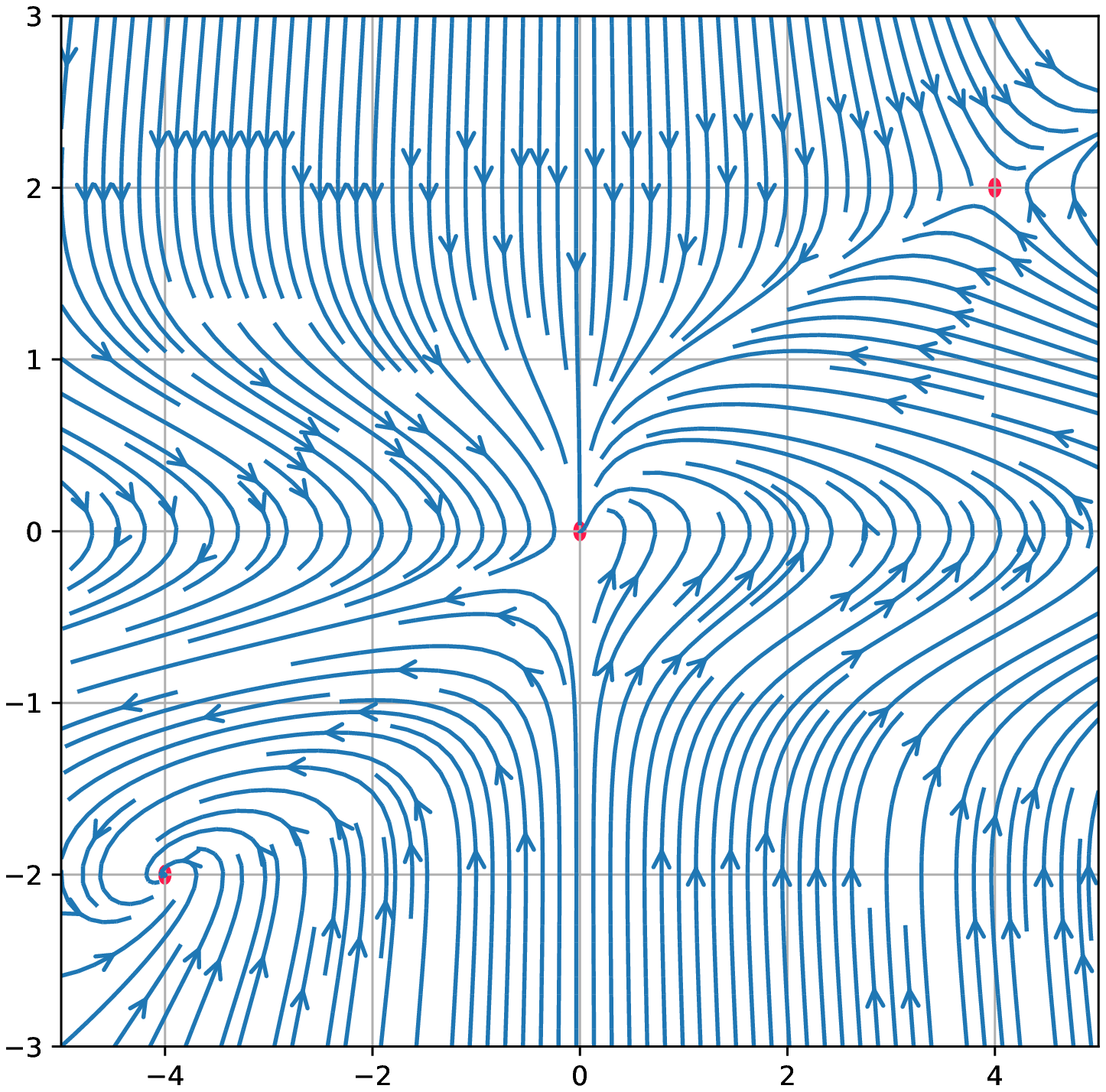}}
  \put(4.8,0){\scriptsize $x_1$}
  \put(1.5,3){\scriptsize $x_2$}
  \put(2.81,1.34){\circle*{0.2}}
  \put(4.87,3.05){\circle*{0.2}}
  \put(6.94,4.75){\circle*{0.2}}
 \end{picture}

 \caption{\textcolor{black}{Phase portrait} associated to the dynamics \eqref{example} with three equilibrium points indicated by the thick dots.}
 \label{Fig:vectorField}
\end{figure}

The system has three equilibrium points given by
\begin{align*}
 x_1=-\frac{c}{b}x_2,\quad x_2 \in\{-a,0,a\}
\end{align*}
with only one being a local attractor and the other two being saddle points (the origin is actually kind of a degenerated saddle point). For the subsequent numerical evaluation the following parameters are considered
\begin{align*}
 a=2,\quad b=3,\quad c=-1.
\end{align*}
The \textcolor{black}{phase portrait} associated to the vector field is generated using the \texttt{streamplot} function of \texttt{matplotlib} in \python ~and is shown in Figure \ref{Fig:vectorField} with the three equilibrium points indicated by thick dots.
Given the complex dynamics of the continuous part and the  lack of observability, even for continuous measurements, the example represents a challenging task for observer design.

In the following the measurement uncertainty is considered as a \textcolor{black}{uniformly} distributed stochastic process truncated on the interval $[-0.1,0.1]$ and is generated in \python ~using the \textcolor{black}{\texttt{random.uniform}} function implemented in the \texttt{numpy} package. The continuous part of the dynamics is solved using the \texttt{dopri5} method implemented in the \texttt{scipy.integrate} package. 

\begin{figure}[!h]
 \begin{picture}(2,6.75)
  \put(0.35,0.25){\includegraphics[scale=0.6]{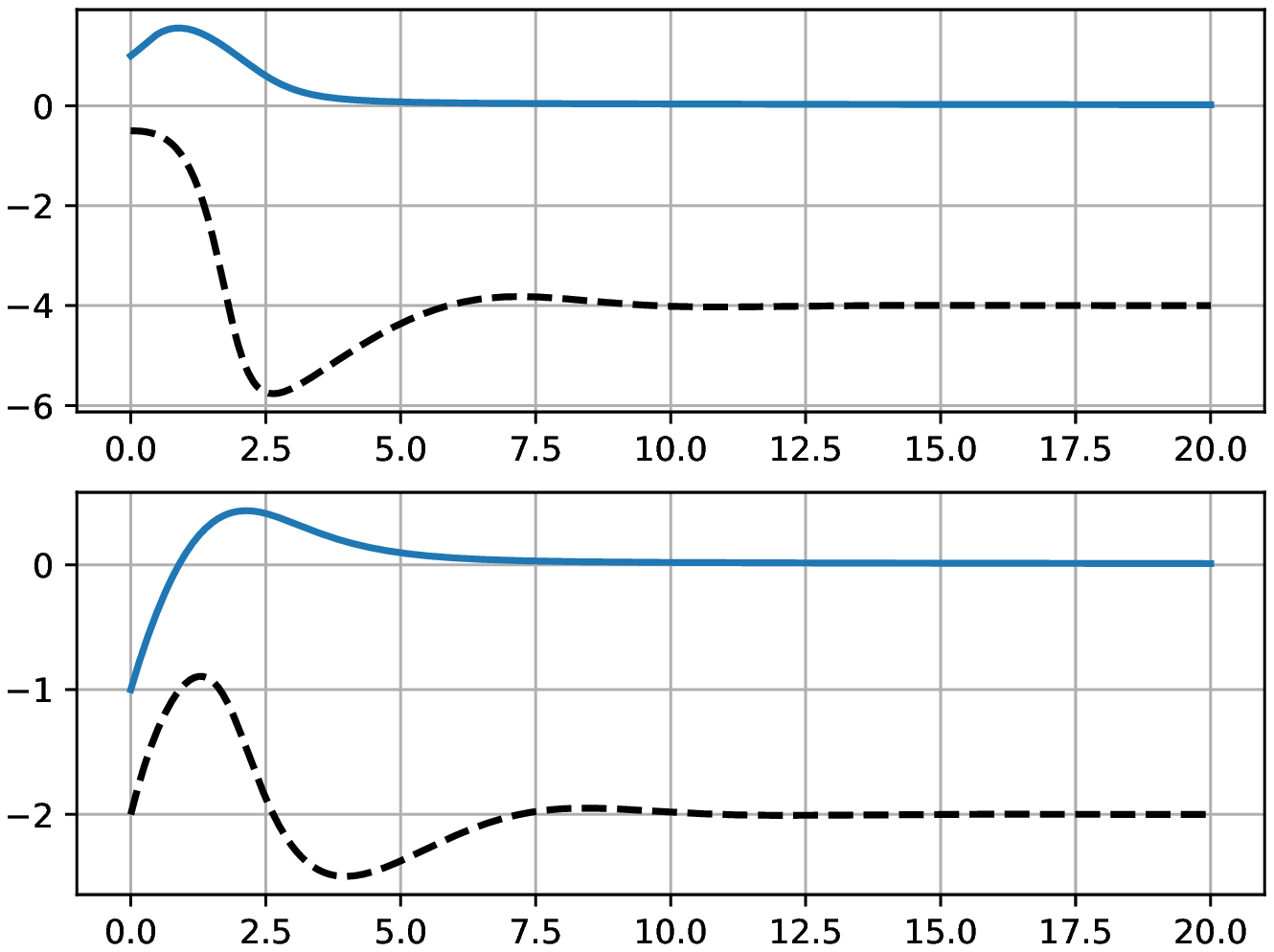}}
  \put(4.4,-0.1){ $t$}
  \put(0,1.9){$x_2$}
  \put(0,5){$x_1$}
 \end{picture}

 \caption{Simulation without innovation by measurement injection.}
 \label{Fig:OL}
\end{figure}

In Figure \ref{Fig:OL} the behavior of a pure simulation without innovation by measurement injection is shown. It can be seen that the observer and system trajectories converge to different equilibrium states and thus the observer does not converge.

\begin{figure}[!h]
 \begin{picture}(2,6.75)
  \put(0.35,0.25){\includegraphics[scale=0.6]{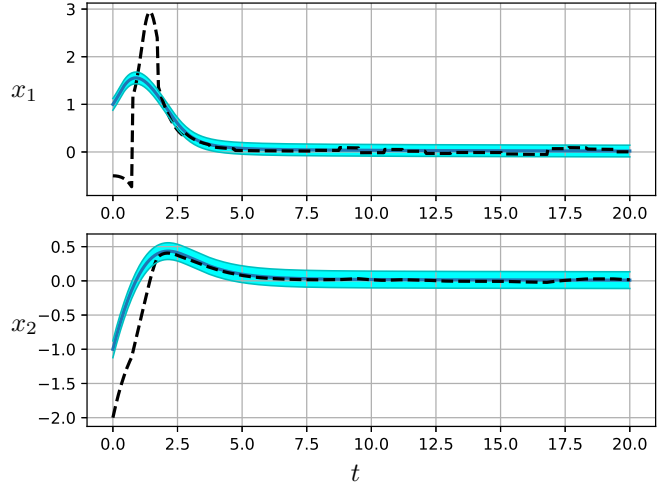}}
  \put(4.4,-0.1){ $t$}
  \put(0,1.9){$x_2$}
  \put(0,5){$x_1$}
 \end{picture}

 \caption{Simulation of the proposed observer (dashed line) in presence of measurement uncertainty. \textcolor{black}{The colored region corresponds to the ISS gain.}}
 \label{Fig:CL}
\end{figure}

\textcolor{black}{For the observer parameters 
\begin{align}
 L = 0.8,\quad T_{min} = 0.63,\quad T_{max} = 1.09.
\end{align}
the conditions of Theorem \ref{thm:ISS_wm} are met with $\alpha=3$. The resulting convergence behavior }
is shown in Figure \ref{Fig:CL}. It can be seen that the observer trajectory (dashed line) converges into \textcolor{black}{the desired} neighborhood of the actual system trajectory.

\begin{figure}[!h]
 \begin{picture}(2,6.75)
  \put(0.35,0.25){\includegraphics[scale=0.6]{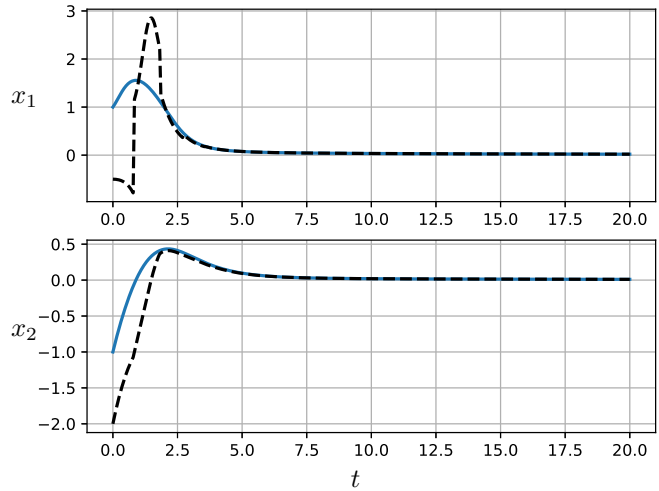}}
  \put(4.4,-0.1){ $t$}
  \put(0,1.9){$x_2$}
  \put(0,5){$x_1$}
 \end{picture}

 \caption{Simulation of the proposed observer (dashed line) in absence of measurement uncertainty.}
 \label{Fig:CL_noNoise}
\end{figure}

Finally, Figure \ref{Fig:CL_noNoise} shows the exponential convergence of the observer in absence of measurement uncertainty.

\section{Conclusions}

The design of a robust nonlinear dissipative observer with impulsive injection of measurement information subject to bounded measurement perturbations is presented. Sufficient conditions on the minimum and maximum time between measurements are derived in terms of system parameters, correction gain and ISS gain by combining notions and tools from impulsive and dissipative systems theory. The effectiveness of the results is illustrated by numerical simulations.

\bibliographystyle{unsrt}

\end{document}

%% file: definitions.tex
\newcommand{\bs}{\boldsymbol}
\newcommand{\bx}{\bs x}
\newcommand{\hbx}{\hat \bx}
\newcommand{\bz}{\bs z}
\newcommand{\hbz}{\hat \bz}

\newcommand{\be}{\bs e}
\newcommand{\beps}{\bs \epsilon}

\newcommand{\by}{\bs y}
\newcommand{\bu}{\bs u}
\newcommand{\bw}{\bs w}

\newcommand{\bpsi}{\bs\psi}

\newcommand{\der}{\mathrm d}
\newcommand{\ddt}[1]{\frac{\der#1}{\der t}}
\newcommand{\dtau}{\mathrm{d}\tau}